\documentstyle[prl,aps,epsfig,multicol,amssymb,amsfonts]{revtex}
\tighten
\renewcommand{\narrowtext} 
{\begin{multicols}{2}\global\columnwidth20.5pc} 
\renewcommand{\widetext}
{\end{multicols}\global\columnwidth42.5pc} 
\begin{document} 
\draft 
\title{Nonadiabatic scattering of a quantum particle in an
inhomogenous magnetic field} 
\author{F.~Dahlem$^1$, F.~Evers$^2$, A.~D.~Mirlin$^{2,3}$,
D.~G.~Polyakov$^2$, P.~W\"olfle$^{2,3}$} 
\address{$^1$Ecole Doctorale de Physique, Universit\'e Joseph Fourier
- Grenoble 1, 38042 Grenoble, France}
\address{$^2$Institut
f\"ur Nanotechnologie, Forschungszentrum Karlsruhe, 76021 Karlsruhe,
Germany}
\address{$^3$Institut f\"ur Theorie der Kondensierten Materie,
Universit\"at Karlsruhe, 76128 Karlsruhe, Germany}
\date{\today}
\maketitle
\begin{abstract}
We investigate the quantum effects, in particular the Landau-level
 quantization, in the scattering of a particle the nonadiabatic classical 
dynamics of which is governed by an adiabatic invariant. 
As a relevant example, we study the scattering
of a drifting particle on a magnetic barrier in the quantum limit where the 
cyclotron energy is much larger than a broadening of the Landau levels induced
by the nonadiabatic transitions. We find that, 
despite the level quantization, the  exponential suppression $\exp(-2\pi d/\delta)$
(barrier width $d$, orbital shift per
cyclotron revolution $\delta$) 
of the root-mean-square transverse displacement experienced by the particle
 after the scattering
is the same in the quantum and the classical regime. 
\end{abstract}

\narrowtext

A well-known example of an adiabatic invariant is the magnetic flux enclosed
by a trajectory of a classical electron drifting in two dimensions in a
magnetic field perpendicular to the drift plane ($E{\times}B$-drift).  
Although it is not a strictly conserved quantity, it implies
strong constraints for the motion of the particle. In particular,
in a smooth potential the the guiding center drifts along
equipotential lines because a shift from the equipotential would change the
cyclotron radius, which violates adiabaticity. This
principle underlies, e.g., the  semiclassical theory of the integer
quantum Hall effect \cite{Trugman83}. It provides a mechanism
for ``adiabatic localization'' of electrons
\cite{Fogler97} and leads to a non-monotonic dependence 
of the width of the cyclotron resonance on the magnetic field
\cite{Fogler98}.

In a series of recent papers \cite{Mirlin98}, it has been emphasized that
adiabatic invariance is crucial also for the
dynamics of composite fermions, which are quasiparticles relevant to
the fractional Quantum Hall effect (FQHE). 
The composite fermions drift along
lines of constant flux in an effective random magnetic field
if the filling fraction sufficiently differs from $1/2$, so that
the average effective magnetic field is strong enough. 
Then almost all trajectories are closed and it is only
due to weak violation of adiabatic invariance that the
conductivity is non-zero \cite{Mirlin98}. 

Because of its significance for the understanding of the
phenomena mentioned above, we have recently formulated and solved
a model problem: the scattering of a classical particle
drifting in a magnetic field with homogenous gradient
towards  a magnetic impurity \cite{Evers98}.
We found that the displacement of the guiding center in a scattering
event is suppressed by an exponentially small factor $\exp(-2\pi
d/\delta)$, where $\delta$ is the shift of the guiding center
in a cyclotron period along the drift trajectory and $d$ is the impurity
size.  In Quantum Hall devices $d$ is determined by the
(spacer) distance between the doping layer and the plane of the
two-dimensional electron gas. 
Moreover, we also found an unusual geometric
resonance  which corresponds to ``self-commensurability". It is
not related to the geometry of the scatterer but to 
that of the unperturbed trajectory: whenever the cyclotron
diameter $2R_c$ and $\delta$ are commensurate
the scattering is strong.

It is an important question whether the strong adiabatic
constraint on the possible classical scattering processes is
lifted by quantum mechanics. If so, then the adiabatic-localization
scenario outlined above and the corresponding interpretation
of experiments could be questionable. 
In order to resolve these issues, we analyze in the present
paper the scattering problem for a drifting particle in the quantum regime.
Like in the classical case, we calculate the root-mean-square (r.m.s.)
transverse shift $\Delta\rho$. In the quantum regime, the Landau-level
quantization is strong, so that a separation in real space of two 
drift trajectories with the same energy and corresponding to different
Landau-levels by far exceeds the classically calculated $\Delta\rho$.
Although one might expect otherwise, we find that the
r.m.s. non-adiabatic shift is given by the same formula as in the
classical case.

An analogous conclusion has been drawn in earlier work for the
case of motion in a random potential \cite{Fogler97}. The difference
between our consideration here and that in Ref. \cite{Fogler97} is
twofold: i) our one-impurity problem has well defined
in- and outgoing states; ii) we consider an inhomogeneous magnetic field
(rather than scalar potential). An important advantage of the
one-impurity scattering problem is that it allows for a 
controlled calculation of the nonadiabatic shift. This yields 
not only the exponential suppression but also the prefactor.
Thus, we find an additional non-trivial
feature, the above mentioned ``self-commensurability" oscillations. 

Let us point out further relations of this  work to yet another field of interest.
Recently, a number of experiments have been dealing with semiconductor
hybrid structures which consist of a quasi one-dimensional conducting
channel with a magnetic barrier that modulates the electronic
current \cite{Johnson97,Kubrak99,Vancura00}.
These experiments have been performed in weak magnetic fields. At 
sufficiently strong fields  the device may be considered an
experimental realization of our model problem.
In principle, this allows for an
experimental check of our predictions  by measuring the
transverse current or the Hall voltage. 
We come back to this issue at the end of the paper. 

The Hamiltonian for the free drift motion of the model problem
formulated above reads 
\begin{equation}
H_0 = \big(i\nabla + {\bf A}_{0}\big)^2\big/2m.
\end{equation}
We choose Landau's gauge with $ {\bf A}_0 = By\big(1+\eta y/2l,0\big)$, where
$l=\sqrt{\hbar/m\omega_c}$ is the magnetic length, 
$\omega_c = eB/mc$ is the cyclotron frequency,
$\eta$ is a dimensionless parameter characterizing the magnetic field gradient
and we adopted units such that $e=c=1$.
Due to translational invariance the
exact eigenstates are of the form
$\psi({\bf x}) = \exp(ikx)\chi_n(y)/\sqrt{L_x}$ 
and the problem is effectively one-dimensional. The functions $\chi_{n,k}(y)$,
with $n=0,1,2\ldots$, are eigenfunctions of a quantum particle in a quartic potential
\begin{equation}
v(y) = \frac{1}{2ml^2}\big[ \frac{y}{l} (1+\frac{\eta}{2l} y) - kl\big]^2.
\label{eq:1}
\end{equation}
After introducing new coordinates $\xi=(y-y_k)/l_k$ one obtains the effective
Hamiltonian 
$$
h_k = \big( -\partial^2_{\xi} + 
(\xi + \eta_k\xi^2/2)^2\big)\big/2ml_k^2,
$$ 
where $l_k=l/\sqrt{1+\eta y_k/l}$, $y_k = \big(\sqrt{1+2\eta kl} -1)l/\eta$ 
and $\eta_k = \eta (l_k/l)^3$. 
The exact eigenfunctions $\chi_{n,k}$ of $h_k$ are not known analytically
and have been calculated numerically\cite{Muller92}.
The simplest approximation is to ignore the anharmonic terms
altogether. 
This procedure is justified if the relative change of the
magnetic field within the cyclotron radius $R_k = \sqrt{2n+1}l_k$ is small: 
$\eta_k R_k/2l_k\ll 1$, which is what we assume here.
In other words, we consider the case where the quartic
potential (\ref{eq:1}) forms a 
double well with two well separated parabolic minima. Our focus is on particle 
energies not too large such that tunneling between the two wells is 
strongly suppressed and the shape of each is approximately parabolic. 
In this situation the electronic trajectories resemble a cyclotron orbit
drifting along lines of constant magnetic field. 

We are then facing a well-defined scattering problem with free scattering
states
\begin{equation}
\chi_{n,k}\approx \phi_{n,k}(\xi) = H_n(\xi)\ \exp\left(-\xi^2/2 \right)\big/N_{n,k}.
\label{eq:3}\end{equation}
Here, $H_n$ denotes the Hermite polynomials and  $N_{n,k}=\big(\sqrt{\pi}2^n
n! \ l_k\big)^{1/2}$ is the normalization factor. 
To leading order in $\eta$ and $1/n$ the spectrum is given by
$
\epsilon_n(y_k) = \hbar \omega_k(n+1/2) 
$
with $\omega_k = \omega_c(1+\eta y_k/l)$.

The magnetic scatterer has a vector 
potential $\delta A_x({\bf x})$ which gives rise to a perturbation 
to the Hamiltonian $H_0$ 
$$
V = \frac{1}{2m}(i\partial_x + A_x(y)) \delta A_x({\bf x})
    + \delta A_x({\bf x}) \frac{1}{2m}(i\partial_x + A_x(y)),
$$                     
where we have dropped the second-order term. 

Because of energy conservation any scattering process from $k$ to $k'$, 
i.~e., from $y_k$ to $y_{k'}$ is accompanied by a simultaneous change 
in the Landau level quantum number $n$. In the parabolic approximation, one has
\begin{equation}
\label{eq:2}
y_k -y_{k'} = (k-k'){{\overline l}^2}= (2\pi {\overline
l}^2/\overline \delta) \, (n'-n),
\end{equation}
where ${\overline l}=l/\sqrt{1+\eta Y/l}$, 
${\overline R_c}={\overline l}\sqrt{(n+n'+1)}$, ${\overline \eta}=\eta
({\overline l}/l)^3$, $Y=(y_k+y_k')/2$ and ${\overline \delta} = \pi
({\overline \eta}/{\overline l}) {\overline R_c}^2$. 
The probability $w(nk;n')$ of a scattering event from an incident state with quantum
numbers $n,k$ to an outgoing state with $n'$ 
can be calculated by means of the golden rule:
$$
w(nk;n') = \sum_{k_f} P(nk;n'k_f)/I_{nk}
$$
with the scattering probability per time 
$P(nk;n'k') = (2\pi/\hbar) \delta(\epsilon_{n,k}-\epsilon_{n'k'})
|M(nk;n'k')|^2$ and the incident current $I_{n,k}=(\hbar
L_x)^{-1}\partial\epsilon_{nk}/\partial k$.  
Within the parabolic approximation one finds
$$
w(nk;n') = |M(nk;n'k')|^2
\left(\frac{\partial \epsilon_{nk}}{\partial k}\frac{\partial
\epsilon_{n'k'}}{\partial k'}\right)^{-1},
$$
where $k'$ is determined by (\ref{eq:2}).
The drift velocity is $\partial\epsilon_{nk}/\partial k{=}
\delta_k\omega_k/2\pi$, where $\delta_k$ 
denotes the shift in $x$-direction of the corresponding classical orbit within
the cyclotron period, $\delta_k {=} \pi (\eta_k/l_k) R_k^2$ in the parabolic approximation. 
The corresponding matrix elements are 
$$
M(nk;n'k') = \langle \chi_{nk}| (\frac{A_x(y)}{m} - \frac{k+k'}{2m})\delta A_x(q,y)|
              \chi_{n'k'}\rangle, 
$$
where $q=k-k'$ and $\delta A_x(q,y)$ is the Fourier transform of the
scattering potential with respect to the $x$-coordinate. 

Before we proceed with evaluating $M(nk,n'k')$ let us 
discuss different relevant regimes.
If the scattering probability $w(nk,n')$ as calculated above from the
golden rule turns out to be large, $w(nk,n')\gg1$, we are in the
quasiclassical regime. In this limit of a strong barrier, when
the Born approximation is not valid, the classical results derived
earlier \cite{Evers98} should be applicable. For a weaker barrier,
$w(nk,n') \ll 1$, the system is in the quantum regime, which is of
central interest for us in the present paper. 

It should be noted that the Born approximation matches the classical
description at $w(nk,n') \sim 1$ directly only if the parameter
$b_0/{\overline \delta}\eta \ll 1$, where $b$ is the characteristic magnetic field
produced by the impurity. If $b_0/{\overline \delta}\eta\ll 1$, one can neglect the
curving of the quasiclassical trajectories by the impurity field.
If $b_0/{\overline \delta}\eta \gg 1$, the Born approximation fails with increasing
$b$ when $w(nk,n')$ is still small, so that there appears an
intermediate regime between the Born and classical limits. In this
regime the scattering probability $w(nk,n')$ can be calculated by means of an
eikonal approximation. The crossover to classics occurs when the exact
scattering probability is of order unity. For our
purpose, however, it is sufficient to consider the limit
$b_0/{\overline \delta}\eta \ll 1$, which we assume in what follows.

We will further assume
the condition $2\pi {\overline l}^2/{\overline \delta} \ll {\overline
R_c}$ (which can be equivalently rewritten as ${\overline \eta}
\sqrt{n^3}\gg 1$ and which ensures that the spacing $\delta y$
between the available states is small compared to the cyclotron
radius). In particular, this condition implies that the Landau
level index is large, $n\gg 1$. In the opposite case ${\overline \eta}
\sqrt{n^3}\ll 1$ (which can be classified as an ultra-quantum limit)
the states with $n'\neq n$ overlap exponentially weakly, which leads
to an additional suppression of the scattering cross-section,
unrelated to adiabaticity. 

Now we return to the evaluation of the matrix element $M(nk,n'k')$.
Since it can be calculated only approximately, we first need to analyze
the impact of gauge invariance. In order to do so, 
we consider a particular example, namely a magnetic field barrier with a shape independent
of $y$, $ \delta A_x(q,y) = (y+y_0)b(q)$. Since the scattering amplitude must
be independent of the choice of the origin $y_0$ we should have 
\begin{equation}
\langle\chi_{nk}| A_x(y) - (k+k')/2|\chi_{n'k'}\rangle = 0
\label{eq:4}
\end{equation}
whenever the states $\chi_{nk}$,$\chi_{n'k'}$ have the same energies.
Indeed, this property of the exact eigenstates follows from the
identity $\langle \chi_{nk}|h_k - h_{k'}|\chi_{n'k'}\rangle =0$, immediately.
Apriori, we cannot expect the parabolic approximation (\ref{eq:3}) to fulfill 
this identity and in fact it does not. To further illustrate this point,  
we introduce the magnetic field $b(q,p)$ corresponding to the scatterer 
by inverting $b(q,y) = -\partial_y \delta A_x(y)$ in Fourier space
\begin{eqnarray}
M(nk;n'k') = && \int \frac{dp}{2\pi} \ b(q,p)/(-ip)  \nonumber \\
             && \langle\chi_{nk}| \left(A_x(y) - (k+k')/2 \right) e^{ipy}|\chi_{n'k'}\rangle 
             \nonumber.
\end{eqnarray}
Due to Eq. (\ref{eq:4}) we can be sure that the pole $p=0$ is 
canceled if the exact matrix element is used. 
In order to avoid spurious terms resulting from 
any approximation that could spoil this cancellation, we employ an exact
rewriting of the formula and replace $\exp(ipy) \rightarrow
\exp(ipy)-1$. 

Even in the parabolic
approximation $\chi_{nk} \approx \phi_{nk}$ the matrix element cannot
be evaluated analytically, since the incoming and 
outgoing states experience slightly different magnetic fields so that
$l_k$ and $l_{k'}$ differ from each other. 
However, the difference may be ignored if the condition
$n,n'\gg |n'-n|$ is met. 
This is precisely the case of interest to  us, since $n,n'\gg 1$
and since the relevant values of $n'-n$ are $\pm 1$, as will be explained
below. 

Therefore we simplify the  matrix element
by letting $l_k \approx l_{k'} \approx {\overline l}$. Dropping further
terms small in ${\overline \eta}, 1/n$ we obtain ($n'\geq n$)
\begin{eqnarray}
M(nk,n'k') &=&  - \frac{1}{2m{\cal N}}\int\!\frac{dp}{2\pi} \ b(q,p)
\ \ e^{ipY} f_{n n'}\big((q+ip){\overline l}/2\big) \nonumber \\
f_{n n'}(\kappa) &=& \kappa^{n'-n} \partial_{|\kappa|^2} \
e^{-|\kappa|^2}L^{n'-n}_{n}(2|\kappa^2|)
\label{eq:5}
\end{eqnarray}
with the normalization factor ${\cal N} = (2^{n'}n!/2^{n}n'!)^{-1/2}$. 
The expression for the other case $n<n'$ is found by interchanging $n',n$ and
replacing $q \leftrightarrow -q$ in $f_{n n'}$.

Let us discuss our result by
considering several relevant examples.
First, we consider the barrier 
$ \delta A_x(q,y) = y b(q)$, again. We will see that higher order transitions
$|n'-n|\geq 2$ are strongly suppressed if the width of the barrier $d$  substantially exceeds
${\overline \delta}$. Since it is precisely the adiabatic limit of smooth
scatterers, $d\gg {\overline \delta}$,
we are mostly interested in, we specialize to the case $n'=n+1$ right away.
Also, we focus on displacements small as compared to the cyclotron radius,
$y_k-y_{k'}< {\overline R_c}$, which allows us to use the asymptotic
expansion of the Laguerre polynomials. 
This yields for the transition probability
\begin{equation}
w(nk; n+1) = \frac{1}{2\pi{\overline l}^2}
\left|\frac{b(q_1)}{m{\overline \omega_c}}\right|^2 \frac{4{\overline
R_c}}{q_1{\overline l}^2} \cos^2\left(q_1{\overline R_c} - \frac{\pi}{4}\right),
\label{eq:6}
\end{equation}
where $q_1=2\pi/{\overline \delta}$. For further discussion, 
we list two examples: relevant to  the
FQHE is the form
$b_{\rm 1}({\bf x}) = b_0 d^3(|{\bf x}|^2 + d^2)^{-3/2}$.
Its one-dimensional Fourier-transform is
$b_{\rm 1} (q,y=0) = b_0 d \sqrt{2\pi |q|d} \exp(-|q|d)$
, valid at $|q|d\gg 1$, which makes the prefactor in the transition probability,
\begin{equation}
w_{\rm 1a}(nk; n+1) = 4 W_0^2 
\frac{{\overline R_c} d^3}{{\overline l}^4} e^{-2q_1 d}
\cos^2\left(q_1{\overline R_c} - \frac{\pi}{4}\right)
\label{eq:7}
\end{equation}
independent of $\eta$ (here $W_0 = b_0/m{\overline \omega}$).
The factor $\exp (-2q_1d)$ represents the
exponential suppression of the transition probability due to
adiabaticity.
For the processes with $|n-n'|>1$ this suppression is much stronger,
$w(nk,n')\propto \exp (-2q_1d|n-n'|)$, so that they can be neglected,
as stated above. 
For the magnetotransport experiments \cite{Kubrak99,Vancura00} 
another magnetic field configuration is relevant, with
an average value equal to zero and two ``bumps" of opposite sign
separated by a distance $a$, i.e. $b(x) = \beta(x-a/2) - \beta(x+a/2)$.
We choose $\beta(x) = b_0 \ln((x^2 + d^2)/(x^2 + (d+D)^2))$ \cite{Peeters98},
which implies
\begin{eqnarray}
w_{\rm 2a}(nk;n+1) &&= \frac{{\overline
R_c}\ {\overline \delta }}{\pi {\overline l}^4} 
\left(\frac{8}{\pi} W_0 \delta \right)^2 
\ \sin^2\left(\frac{q_1 a}{2}\right) e^{-2 q_1 d} \nonumber \\ 
&& \times \sinh^2(q_1 D)\ 
\cos^2\left(q_1{\overline R_c} - \frac{\pi}{4}\right)e^{-q_1 D}.
\end{eqnarray}
In addition to the adiabatic suppression, we observe an 
oscillating behavior as the slope ${\overline \eta}$
is varied, with a geometrical resonance condition
$a=j_o {\overline \delta}, j_o=\pm1,\pm2,\ldots$
at which the leading order scattering process is forbidden. 

Next, we turn to another example where the scatterer is of finite extent
also in $y$-direction.
We evaluate the integral for the case of $b_{\rm 1}$,
as defined above, in two different limits, again focussing on $n'=n+1$. 
When ${\overline R_c}$
exceeds all length scales, $d,Y\ll {\overline R_c}$, we obtain
\begin{equation}
w_{\rm 1b}(nk;n')=4 W_0^2 (d/{\overline l})^4 \cos^2(q_1{\overline R_c}) e^{-2q_1 d}
\end{equation}
for the transition probability. In the other regime, where
${\overline R_c}\ll d$ and $Y=0$, we recover the earlier result
Eq. (\ref{eq:7}). 

We are in a position now to compare the classical and quantum
mechanical results for the quantity $\Delta \rho$ which is defined as 
the r.m.s. magnitude of the shift perpendicular to drift
velocity after the scattering took place, 
$\Delta\rho=\langle (\delta y)^2\rangle^{1/2} = (2\pi{\overline l}^2/{\overline \delta})(\sum_{n'} (n-n')^2 w_{nk,n'})^{1/2}$.
For the case ${\overline R_c}\ll d$ and $Y=0$ we have
\begin{equation}
\Delta\rho_{\rm 1a} = 2^{5/2} \pi \, W_0 \, 
\sqrt{{\overline R_c}d^3/{\overline \delta}^2} \ e^{-q_1 d} \cos\left(q_1
{\overline R_c}{-}\pi/4\right),
\end{equation}
whereas when  $d,Y\ll {\overline R_c}$  we find
\begin{equation}
\Delta\rho_{\rm 1b} = 2^{5/2} \pi\, W_0 \, (d^2/{\overline \delta}) \ 
e^{-q_1 d} \cos\left(q_1 {\overline R_c}\right)
\label{eq:11}
\end{equation}
under the conditions ${\overline \eta}\sqrt{n} \ll 1\ll {\overline
 \eta} \sqrt{n^3}$ and $d/{\overline \delta}\gg 1$. 
These results are identical
with the classical findings reported in Ref. \cite{Evers98}
after the latter have been averaged over the initial conditions.
Note, however, that the picture of the scattering is totally different
in the classical and the quantum regimes. In the former case, it is
 the typical magnitude of the adiabatic shift which is exponentially
 small $\delta y \propto \exp(-2\pi d/{\overline \delta})$. On the
other hand, in the quantum case the scattering shift is set by the
 spacing between the unperturbed states, $\delta y = 2\pi {\overline
 l}^2/{\overline \delta}$, while the scattering probability is
exponentially suppressed $w \propto \exp(-4\pi d/{\overline \delta})$.
So, the agreement found between the results for the r.m.s. shift is
far from trivial. It is also worth mentioning that it is exactly the
 r.m.s. shift that  determines the macroscopic diffusion coefficient
in a system with a finite density of scatterers \cite{Fogler97,Mirlin98}.

It is known that in the absence of a magnetic field the transport
scattering rate has the same value in the classical and quantum
regimes, provided the correlation length of the randomness is
large, $k_F d \gg 1$ \cite{Dyakonov91}. We find now an analogous result
for the case of transport in strong magnetic fields, when the dynamics
is governed by the adiabatic invariance. 

Next, we briefly discuss scattering events in which the displacement
$y_k-y_{k'}=q{\overline l}^2$ exceeds the cyclotron radius,
$q{\overline l}^2 > {\overline R_c}$ (``ultra-quantum limit").
In other words, we focus on very small gradients ${\overline \eta}\sqrt{n^3}<1$. 
Once again we consider the magnetic barrier independent of $y$. 
Then the scattering probability is 
\begin{eqnarray}
w(nk,n') = && \frac{1}{2\pi{\overline l}^2} \left|\frac{q_1{\overline l}^2} {{\overline R_c}} 
 \ \ \frac{b(q)}{m{\overline \omega}} \right|^2 \left( \frac{q{\overline l}^2}{{\overline R_c}}\right)^{2(n'+n)} 
  \nonumber \\
 && \times e^{-(q^2{\overline l}^2 - {\overline R_c}^2/{\overline l}^2)},
\label{eq:13}
\end{eqnarray}
provided that $|n'-n|\ll n,n'$.
In the typical case, where the magnetic scattering field has a Fourier transform $b(q) \propto e^{-qd}$, 
the Gaussian factor $\exp(-q^2{\overline l}^2)$ reflecting the shape of
the eigenfunctions starts to dominate over the adiabatic factor
$\exp(-2qd)$ at $q{\overline l}^2 \approx {\rm max}(2d,{\overline R_c})$. 

Up to now we have considered the motion of a particle drifting in a
magnetic field with  homogenous gradient. The same analysis can be
done also for particles that perform an $E{\times}B$-drift. 
The transition probability is now
$
w(nk;n') = |M(nk;n'k')|^2/v_d^2
$
where we have chosen the $E-$field ${\bf E}=(E_x,0)$ and
introduced the corresponding drift velocity $v_d=E_x/m\omega_c$.
The matrix element is given by Eq. (\ref{eq:5}).
With the substitutions $\delta = 2\pi v_d/\omega_c$ and $\eta=0$
the results (\ref{eq:6})-(\ref{eq:13}) derived thereof remain valid. 

We have mentioned already that in general the scattering to the left and
to the right have different probabilities. In the case
of the $E{\times}B$-drift the transverse current resulting from
processes with $n'=n\pm 1$ can be easily evaluated: 
$$
I_y = n_{\rm e} \omega_c\left( b(q)\big/ m\omega_c \right)^2
\cos (2 q R_c)/\pi ,
$$
where $n_{\rm e}$ denotes the particle density. 
Similar to the case of $\Delta\rho$ the
current becomes very small whenever the condition $d\gg \delta$ is
satisfied. Therefore we predict a dramatic decrease of the transverse current
upon crossing over into the regime of adiabatic dynamics at the
barrier. 

Finally, let us mention that the calculation above
implies a constant density of states.
However, for electrons in smooth disorder the density of states splits
into separate Landau bands for sufficiently large $B$. It remains to
be seen what will be the implications of adiabaticity if the crossover
to drift occurs at still larger $B$. We expect a pronounced signature
of the crossover to adiabaticity in the amplitude and shape of the
magnetoresistance oscillations.

It is a pleasure to thank V. Kubrak, B.~L. Gallagher and I.~V.~Gornyi
for interesting discussions. 
Financial support from SFB 195 of Deutsche Forschungsgemeinschaft,
of the DFG-Schwerpunktprogramm ``Quanten-Hall-Systeme" and from INTAS
grant 97-1342 is gratefully acknowledged.

\end{multicols}
\end{document}